\documentclass[11pt]{article}

\usepackage{epsfig}
\usepackage{psfrag}
\usepackage{epsf}
\usepackage{amsmath}
\usepackage{amssymb}
\usepackage{fullpage}
\usepackage{graphicx}
\usepackage{verbatim,moreverb,alltt}
\usepackage{fancyvrb}
\usepackage{verbatimbox}
\usepackage{scilab}

\title{A novel code generation methodology for block diagram modeler and simulators Scicos and VSS}
\author{Jean-Philippe Chancelier\thanks{This code generation tool was developed during a three-years research 
projects funded within the French FUI 2011 called ``Projet P''~\cite{P-project}}\\Université Paris-Est, Cermics (ENPC) \and Ramine Nikoukhah\footnotemark[1]\\Altair Engineering}

\def\scicosscale{0.5}
\def\pictures{}
\def\mybox#1{#1}

\makeatletter
\newcommand{\verbatimfont}[1]{\renewcommand{\verbatim@font}{\ttfamily#1}}
\makeatother

\begin{document}
\maketitle

\begin{abstract}
Block operations during simulation in Scicos and VSS environments 
can naturally be described as Nsp functions. But
the direct use of Nsp functions for simulation leads to poor performance since the Nsp language
is interpreted, not compiled. The methodology presented in this paper is used
to develop a tool for generating efficient compilable code, such as C and ADA, for Scicos and
VSS models from these block Nsp functions. Operator overloading and partial evaluation are the 
key elements of this novel approach.
This methodology may be used in other simulation environments such as Matlab/Simulink.
\end{abstract}

\tableofcontents

\section{Introduction}

In this paper we present a new methodology of code generation for block diagram simulation tools 
such as Scicos~\cite{scn}, VisSim SIMULATE\footnote{VisSim SIMULATE is a commercial modeling and simulation product
based on Scicos.} (VSS) and Simulink. 
This methodology takes advantage of the operator overloading facilities
available in matrix based languages such as Matlab, Scilab, Nsp~\cite{nsp} or Octave. The implementation is 
done in Nsp/Scicos and VSS environments \cite{scn} but the code may easily be ported to 
Matlab/Simulink and other
similar environments. In our implementation, the final C code can be either 
generated directly using a C pretty printer or
through the use of Gnat Model Compiler \cite{gnat}. In the latter case, an \verb+.xmi+ 
file is generated and then
\verb+gmc+ is invoked to generate the C code. Either way, the C code may be 
reimported automatically in the Scicos and VSS environments as a \verb+CBlock+ for validation.

Not only the block behaviors may easily and naturally be implemented in the Nsp language but
they can also be simulated directly in Scicos, if implemented as such. The implementation is natural
because most Scicos blocks support matrices and Nsp data types. But also because Nsp is the working
language in the Nsp/Scicos environment. In particular Scicos model and block parameters are defined via Nsp 
scripts and expressions, and evaluated by the Nsp interpreter. 

The main reason for which only parameter definitions in Scicos blocks use Nsp programs and run-time block
code is invariably expressed in C is performance. Parameters are evaluated once at compile time, 
so peformance is not an issue, but if the run time block behavior is expressed in Nsp, the simulation 
performance would be significantly reduced.

The methodology proposed in this paper provides the possibility of defining Scicos/VSS blocks entirely
in Nsp. These blocks, which can be developed by Scicos developers, toolbox developers or end-users,
may be tested and validated directly by (low performance because interpreted) 
simulations in Scicos. The Nsp code of the
block can then be used for code generation by the methodology presented in this paper. 

In this methodology, an Nsp script is generated from the Scicos/VSS model. The generation of this script 
is based on the result of the compilation of the model by Scicos or VSS compilers. The compiler result 
includes in particular the order in 
which the blocks in the model must be executed (their
functions called) and the types and sizes of all the block inputs and outputs. This generated script,
which contains calls to the Nsp functions of the blocks, evaluates block outputs, new discrete states,
state derivatives, zero-crossing surfaces, etc., i.e., all the information needed for both simulation
and code generation.  

The main obstacle in generating compilable code (such as C or ADA) from Nsp, and more generally from
all Matlab-like languages, is the inability, in the general case, 
to determine the types and sizes of variables before run-time.
But this is not an issue for the script generated from a Scicos or VSS model since in this case, 
all the sizes and types of the variables representing block inputs and outputs are known 
(determined by the compiler). 
The main component of
the code generation tool presented here is the Nsp code generation facility that can produce
compilable code from such Nsp scripts. It should be emphasized that this facility is not a general
purpose code generator for arbitray Nsp scripts. It is developed especially for Scicos/VSS usage where
in the generated code, data types involved are limited to matrices of doubles and various integer 
data types, and all variable sizes can be statically determined.

This Nsp code generation facility is entirely developed in Nsp using its abilities to define 
new data types and performing operator overloading for these new data types.
In particular we define a new data type called \verb!bvar! representing a matrix and containing:
\begin{itemize}
\item type: numerics or symbolics
\item value: a matrix
\item name: string
\end{itemize}
If a variable of type \verb!bvar! is of type numerics, then its value represents its actual value
and it is similar to a regular Nsp matrix with the same value. However using the type \verb!bvar!, it
is possible to associate a name to the variable. If the variable is of type symbolics, then
its value is just a ``nominal value'', often zero, but its data type (double, int8, int16,...) 
and its size represent the actual data type and size of the variable. 

Operations involving numeric variables (of type \verb!bvar!) and regular Nsp variables never produce
symbolic variables (of type \verb!bvar!). Operations involving symbolic variables produce in general
symbolic variables except in special cases. In particular if \verb!X! is a symbolic variable, then
\verb!size(X)! is not; neither is \verb!datatype(X)!. The overloading of basic operators and language
primitives for these data types automatically overloads all Nsp functions that use these operators and
primitives thanks to dynamic scoping property of the Nsp language.

The main idea of the approach used in this paper is to overload  basic Nsp operators 
and functions for variables of type \verb!bvar! so that if we have an Nsp script that performs
operations on regular matrices, if these matrices are replaced by variables of  type \verb!bvar!,
the script runs as before but in addition, a pseudo-code (a sequence of ``instructions'')
is generated and stored in a global list variable named \verb!code!.
Every time an operation involves a variable of type \verb!bvar!, the overloaded
operation performs the original operation returning the corresponding value as if the arguments
were not of type \verb!bvar! but in addition it adds one or several ``instructions'' to
the list \verb!code!. This mechanism may be seen as spying the script
as it is executed: when an operation involving a type \verb!bvar! variable is performed, it
is recorded in the \verb!code! list. The record includes not only the nature of the operation (for
example matrix multiplication) but also certain information on its operands. The information in
the generated pseudo-code is rich enough so that it can be used to generate code in C, ADA or other
languages, by simple pretty printing operations.

 The execution of the script  can be seen
as partial evaluation. When the value is unknown, only the type and size information is propagated;
when the value is known, the actual value is propagated. As long as the script can run to the end by
propagating this partial information, the code generation can be performed.

Unlike most code generators where the objective is to 
transform a program written in a source language into a program in the target language preserving its
functionality, in our approach the program in the source language contains not only the code to
be translated but also specifications on how this transformation should be performed. This is done
by providing a set of functions for use as
code generation directives so that a code generation task may be expressed as a part of the Nsp script.
The Scicos/VSS block Nsp codes however will not make use of these functions. 
Scicos users need not know about these directives; as far as they are concerned, 
the block codes are written in standard Nsp language with regular Nsp data. During code generation,
the script will be calling these block codes with the variables expressing blocks input/outputs
and states defined as symbolic \verb!bvar! variables unbeknownst to the Scicos user.

There are many advantages in using the technique presented here as opposed to developing an independent
code generator: this code generator does not use a different parser for the scripting language; the script 
is run using the tool itself. Similarly, the operator and functions used in the
process of partial evaluation are that of the scripting language itself. 
Finally the operator overloading is being
performed in the scripting language, so it can be customized easily for different targets. 

The method presented here 
is implemented in Nsp but it may be developed in other languages providing similar overloading facilities, 
in particular Matlab, Scilab and Octave.

The tool based
on this methodology provides a powerful and easily customizable code generator for Scicos and VSS. 
Despite the
existence of code generators for Simulink (Matlab and Simulink coder in particular) a 
tool based on the approach presented in this paper  would  still be of interest since it provides 
more control over the generated code.

In this paper we will focus on code generation for discrete-time subsystems of Scicos and VSS 
models. This is what is needed in the  majority of embedded code generation applications, in 
particular for the implementation
of embedded controllers. We consider double, boolean and signed and unsigned 8, 16 and 32 bit integers
data types.
We do not consider the complex data type (seldom used in embedded code), nor do we consider fixed-point
data types, which will be considered in a subsequent paper and for which the methodology presented here is
very promising.

\section{Simple Example}
\begin{figure}[ht]
  \begin{center}
    \mybox{\includegraphics[scale=\scicosscale]{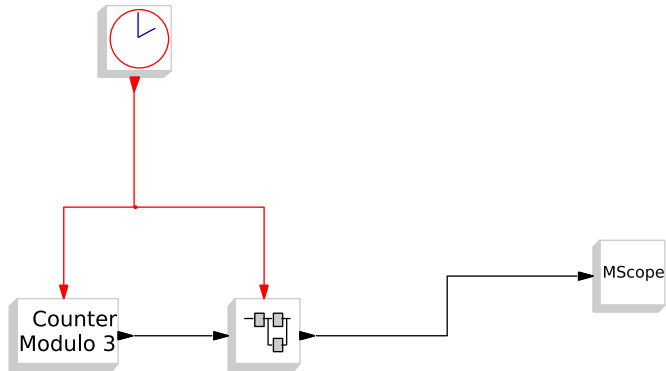}}
    \caption{Main diagram.\label{m1}}
  \end{center}
\end{figure}

\begin{figure}[ht]
  \begin{center}
    \mybox{\includegraphics[scale=\scicosscale]{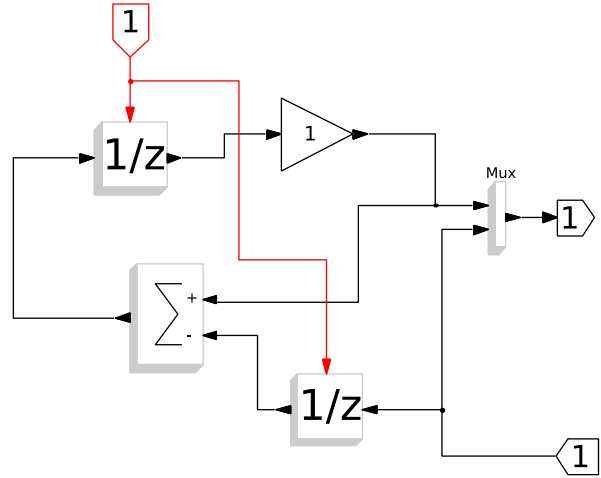}}
    \caption{Content of Super Block for which C code is generated.\label{m2}}
  \end{center}
\end{figure}

We start with a very simple example to illustrate the basic idea of our approach. Consider the
Scicos model depicted in Fig.~\ref{m1}. 
The objective is to generate C code for the Super Block the contents
of which is depicted in Fig.~\ref{m2}.
 

Scicos code generator produces an Nsp script, based on the result of the Scicos compiler,
corresponding to the subsystem inside the Super Block. This script calls the Nsp functions associated
with the blocks in the subsystem. The block functions have an imposed calling sequence and
provide specific callback functionalities. Here we review the functions associated
with the basic blocks used in this subsystem to illustrate how block functions may be developed for
code generation.

A simplified version of the Nsp code of the delay block \verb!1/z! is given below:
\begin{verbatim}
function block=P_Unit_Delay(block,flag)
  if flag==1 then
     block.io(2)=block.state(1)
  elseif flag==2 then
     block.state(1)=block.io(1)
  elseif flag==-1 then
     z=convert(block.params.p1,datatype(block.io(1)))
     if prod(size(z))==1 then z=z*convert(ones(size(block.io(1))),datatype(z));end
     block.state(1)=z
  end
endfunction
\end{verbatim}

The arguments of the block functions are an Nsp structure and an integer. 
The structure, named \verb!block! in this code, contains the fields
\begin{itemize}
\item \verb!io!: a list of matrices of possibly different types and sizes representing block input and output values
\item \verb!state!: a list of matrices of possibly different types and sizes representing block states
\item  \verb!params!: an Nsp structure coding the names and values of block parameters.
\end{itemize}
The second argument of the function is a flag that indicates what job the function should perform. The
value $-1$ is used for initialization. The function is called with this flag once at the beginning so that
it can initialize the internal states of the block, if any. Flag values $1$ and $2$ are used 
respectively to code the output update and state update phases of the block behavior. Here,
we consider only code generation for discrete-time subsystems. For more general blocks, other
flag values are used for example for the computation of state derivative, evaluation of zero-crossing
surfaces, etc. For more information on the usage of flags see \cite{scn}, Chapter~9.
 
In the script generated from the Scicos or VSS model, 
the Nsp code of the blocks are called several times with different flag values.
Unlike block parameters, which are known at the time of
code generation, block input/output and state values are not known. Their values are defined as 
\verb!bvar! variables when the block function is called. Even though the function has been developed
to receive a structure containting regular matrix arguments, thanks to operator overloading, 
the function accepts this new argument.

The structures \verb!io! and \verb!state! are also new data types for which insertion and
extraction operations have been overloaded. But once again, the developer of the block function code
needs not know about it.

When this function is called with flag $-1$, normally we would simply copy the parameter value in the
state, but the \verb!Delay! block accepts that the parameter be defined as a scalar, even if the
state is not scalar, in which case all the entries of the state matrix take the value of the parameter. Moreover,
even if the parameter is defined as a double (default data type in Nsp), the block may be of a
different data type so the parameter should be converted into
correct data type before being copied in the state. Note that the expressions
\verb!datatype(block.io(1))! and \verb!size(block.io(1))! in this code can be evaluated even if
\verb!block.io(1)! is symbolic (their values depend only on the data type and size of \verb!block.io(1)!). 
Needless to say that 
the execution of this part of the code does not perform any code generation, and thus no trace
of this code will be seen in the final generated code.

For this block, the state update and output update operations (Flag values $1$ and $2$) consist
simply of copying respectively the input to the state, and the state to the output. The corresponding
pseudo-code is generated thanks to the overloading of the insertion operation in 
the \verb!io! and \verb!state! fields.

The simplified version of the code for the Gain block is as follows:
\begin{verbatim}
function block=P_GAINBLK(block,flag)
  if flag==1 then
    put_annotation("Gain block begins.")
    block.io(2)=convert(block.params.p1,datatype(block.io(1)))*block.io(1)
    put_annotation("Gain block ends.")
  end
endfunction
\end{verbatim}
This block does not have a state, so its function contains no initialization phase or state update phase.
As in the case of the previous block, the block parameter, the gain value in this case, may be defined
as a double even if the block data type is something else, for example any type of integer. 

The Gain block in Scicos may perform different types of operations. The gain parameter may be a matrix
and the block output is the product of this matrix with the input vector (which must have compatible sizes).
But the gain parameter may also be a scalar, in which case the input and output have identical sizes
and the product multiplies all the elements of the input matrix with this parameter to compute the output.
But the parameter may also be a matrix when the input is a scalar. But all of this complexity is captured
in the Nsp multiplication operation used in the code. Indeed the Nsp multiplication operation has
exactly the same properties. This similarity is often seen between basic Nsp operators and primitives, and
basic Scicos and VSS block behaviors. This is the fundamental reason why
using Nsp coding is a convenient way of describing the 
behavior of Scicos and VSS blocks, often leading to a compact, readable and maintainable Nsp code.

The function \verb!put_annotation! is used to place comments in the generated code.

We now give the  complete code of the \verb!SUMMATION! block function because it shows some advanced
features of  block definitions that can directly be handled by our approach:
\begin{verbatim}
function block=P_SUMMATION(block,flag)
 if flag==1 then
  vars=block.io
  nin=length(vars)-1
  sgns=block.params.p2
  put_annotation("Sum block begins with "+string(nin)+" inputs.")
  if nin == 1 then
     put_annotation("Using the sum function.")
     if sgns(1)==-1 then
       out=-sum(vars(1))
     elseif sgns(1)==+1 then
       out=sum(vars(1))
     else
       error('wrong sign: "+string(sgns(1)))
     end
  else
     if sgns(1)==-1 then
       out=-vars(1)
     elseif sgns(1)==+1 then
       out=vars(1)
     else
       error('wrong sign: "+string(sgns(1)))
     end
     for i=2:nin
       if sgns(i)==-1 then
         out=out-vars(i)
       elseif sgns(i)==+1 then
         out=out+vars(i)
       else
         error('wrong sign: "+string(sgns(i)))
       end
     end
  end
  block.io($)=out  // $ indicates the last element of the list
 end
endfunction
\end{verbatim}
This block performs two very different summing operations. When the block has only one input, the
output of the block is the sum of the entries of the input matrix. When it has more than one inputs,
the block performs the addition (or substraction depending on the value of parameter \verb!p2!) of the
input matrices. The Nsp code contains both algorithms and the selection is made by testing the
number of inputs. This test and other tests in particular for deciding if the operation to perform is
an addition or a subtraction are ``partial evaluated'' and will not appear in the generated code.

Note that the conditional expressions of the \verb!if! statements and the range of \verb!for! loops 
must be known (through partial evaluation) at the time of code generation. For example in this case,
\verb!nin!, which represents the number of inputs, is known because the length of \verb!block.io! is known.
The generated code will of course contain no trace of these conditional statements or loops.
Clearly if the conditional expression of an \verb!if! statement is not known, the execution of the
script cannot continue. This is a limitation of this approach. 

Note also that the \verb!error! functions generate errors at the time of code generation if a parameter
value is incorrect. But if the Nsp code is used as run time code for simulation, they generate 
error messages at the beginning of the simulation if a parameter does not have correct value. 

The Nsp code for the Scicos \verb!MUX! block is also straightforward since the operation of this block
corresponds to row concatenation in Nsp. The block may have more than two inputs in which case repeated
concatenations are required within a \verb!for! loop:
\begin{verbatim}
function block=P_MUX(block,flag)
    if flag==1 then
      vars=block.io
      y=vars(1)
      put_annotation("MUX block begins with "+string(length(vars)-1)+" inputs.")
      for i=2:length(vars)-1
         y=[y;vars(i)]
      end
      block.io($)=y
      put_annotation("MUX block ends.")
    end
endfunction
\end{verbatim}

The generated C code for the Super Block in Fig.~\ref{m2} contains 
two functions for state update and output update:
\begin{verbatim}
#include <scicos/scicos_block4.h>
#include <string.h>
#include <stdio.h>
#include <stdlib.h>
#include <stdint.h>
#include <math.h>
typedef int boolean;
/* Start1000*/

static double  z_10001=0;
static double  z_10002=0;
static double  link10004=0;

void initialize1000(){
static double  tmp_13=0;
static double  tmp_14=0;
static double  tmp_15=0;
   z_10001=tmp_13;
   z_10002=tmp_14;
   link10004=tmp_15;
}

void updateOutput10001(double *inouts1,double *inouts2){
double tmp_1;
double tmp_5[2];
   /* Gain block begins.*/
   /* Gain block ends.*/
   tmp_1=z_10001;
   /* Sum block begins with 2 inputs.*/
   link10004=(tmp_1-z_10002);
   /* MUX block begins with 2 inputs.*/
   tmp_5[0]=tmp_1;
   tmp_5[1]=*inouts1;
   /* MUX block ends.*/
   inouts2[0]=tmp_5[0];
   inouts2[1]=tmp_5[1];
}

void updateState10001(double *inouts1,double *inouts2){
   z_10001=link10004;
   z_10002=*inouts1;
}

/* End1000*/

void toto1000(scicos_block *block,int flag)
  {
  if (flag == 1) {
   updateOutput10001((GetRealInPortPtrs(block,1)),(GetRealOutPortPtrs(block,1)));
  }
  else if (flag == 2) {
   updateState10001((GetRealInPortPtrs(block,1)),(GetRealOutPortPtrs(block,1)));
  }
  else if (flag == 4) {
     initialize1000();
  }
\end{verbatim}

The calling sequence of the state update and output update C routines of the block contain
pointers to the input and output data, in that order. The ``state'' of the block, which
consists of the subsystem blocks' input/outputs (link values), and the state of the Delay block
 are defined as global variables so that their values
are preserved from the output update phase to state update phase, and from one time step to the next.
Not all link values need to be treated as global, further code optimization may be used to identify
link values that need not be preserved and thus simplify the
code by removing these links out of the global variables. 
In this case additional optimization will also
remove the unnecessary temporary variables \verb!tmp_1! and \verb!tmp_2!, but the C compiler does
this anyway.

Even though there is no scalar type in Nsp (a scalar is just a one by one matrix) 
the generated C code uses, for example for real values, the
type \verb!double! for scalar variables and not \verb!double *!, which is used for non-scalars only. But
the arguments of function calls are all passed as pointers.

\begin{figure}[ht]
  \begin{center}
    \mybox{\includegraphics[scale=\scicosscale]{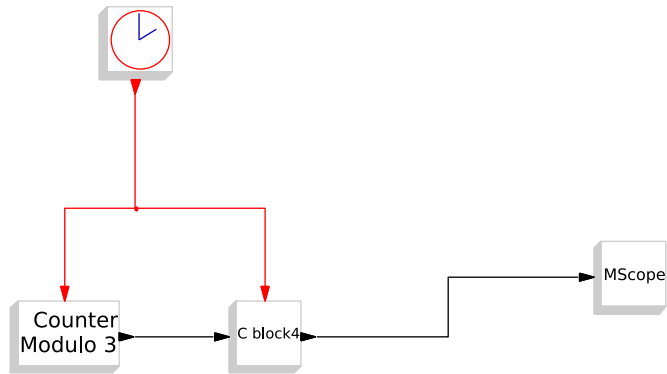}}
    \caption{Super Block is automatically replaced with a C block for testing.\label{m4}}
  \end{center}
\end{figure}

Following the code generation process, the Super Block is automatically replaced in the model
with a \verb!CBlock! containing
the generated C code; see Fig.~\ref{m4}. The simulation of this new model can be used to 
compare simulation results before and after code generation. This is a fast and efficient method 
for testing the generated code.

\section{Operator overloading and pseudo-code generation}
Overloading of operators and primitives is a powerful method in Matlab like environments used
for various applications, such as automatic differentiation~\cite{wr,pwr}. Overloading can be used
to alter the behavior of an existing code, in particular by adding additional operations such as
the evaluation of the derivatives for automatic differentiation, or pseudo-code generation in our case.

\subsection{New data types and overloading}
In Nsp~\cite{nsp}, defining new datatypes for which operator overloading is possible is done by 
writing the new datatype definition in a C code. Some Nsp internal tools 
are available to ease the writing of the code necessary to implement a new datatype definition. 
As an example, the C-code needed to implement the \verb+bvar+ datatype was generated from the 
following simple definition of the \verb+bvar+ object.

\begin{verbatim}
(define-object Bvar
  (in-module "Bvar")
  (parent "Object")
  (c-name "NspBvar")
  (fields
   '("gboolean" "sym" "hidden" "FALSE"); 
   '("NspObject*" "value" "hidden" "NULL" ); 
   '("char*" "varname" "hidden" );
   )
  (gtype-id "Bvar")
)
\end{verbatim}

When the new datatype is defined, the definition of the overloaded operators can be performed in Nsp or in C.
Examples on how to define new data types in Nsp are given in the \verb!src/types-test! directory of 
Nsp source code. 

The main new data type used in this application is called \verb!bvar! (block variable). 
An object of type \verb!bvar! is a wrapper around a standard Nsp object decorated with new attributes: 
an attribute containing the name of the variable and an attribute to specify whether the object is 
numeric or symbolic.
The following lines of code explain how to create a \verb!bvar! variable, which encapsulates a 
\verb!1x1! symbolic matrix of type double named \verb!x! having nominal value~67.

\begin{verbatim}
-nsp->A=bvar(varname="x",value=67,symbolic=%t)
A	= "x",%t,Mat
-nsp->type(A,'short')
ans	=		s (1x1)

  bvar
-nsp->A.get_value[]
ans	=		r (1x1)

 |  67 |
-nsp->A.get_varname[]
ans	=		s (1x1)

  x
-nsp->A.is_symbolic[]
ans	=		 b (1x1)

 | T |
\end{verbatim}

The fields in the \verb!A! variable can be obtained or changed using methods, and the type 
of object \verb!A! is obtained by using the \verb!type! function. 
The short type name of a variable is the string which is used to define functions used 
for overloading. Suppose we define the function \verb!f_bvar! in a function library as follows~:
\begin{verbatim}
function y=f_bvar(A) y=size(A.get_value[]);endfunction;
\end{verbatim}
Then calling \verb!f(A)! will return the size of the value stored in the \verb!bvar! variable \verb!A!:

\begin{verbatim}
-nsp->f(A)
ans	=		r (1x2)

 |  3  4 |
\end{verbatim}

Nsp operators and functions used for writing Scicos and VSS blocks are overloaded for \verb!bvar! 
variables. As an example we give the code 
for the function \verb!inv_bvar!, which implements matrix inversion.
When the matrix to be inverted is \verb!2x2! we use the explicit inverse formula. 

\begin{verbatim}
function out=inv_bvar(in)
   [m,n]=size(in);
   if m<>n then error("Division by non square matrix not supported.");end
   if m==1 then 
     out=1/in
   elseif m==2 then
    out=bvarempty(in)
    out(1,1)=in(2,2)
    out(2,2)=in(1,1)
    out(1,2)=-in(1,2)
    out(2,1)=-in(2,1)
    out=out/(in(1,1)*in(2,2)-in(1,2)*in(2,1))
   else
    .... 
   end
endfunction
\end{verbatim} 

When used with a \verb!2x2! symbolic matrix, the evaluation of the above 
function executes the previous code, which emits corresponding pseudo-code directives stored in 
a global variable called \verb!code!. This variable can then be used to produce 
code for different targets. Here is the C code obtained after calling 
\verb+inv+ on a symbolic \verb!2x2! matrix:

\begin{sessioncmd}
-nsp->codegen_init();
-nsp->A=symbolics(rand(2,2));
-nsp->inv(A);
-nsp->global('code');
-nsp->[L,code,declarations]=code_optimize(list(),code,list(),list(),opt=[
-nsp->txt=code_printer_c(code,list())
txt     =               s (10x1)

    tmp_2[0]=(tmp_1[3]);                                     
    tmp_2[3]=(tmp_1[0]);                                     
    tmp_2[2]=(-(tmp_1[2]));                                  
    tmp_2[1]=(-(tmp_1[1]));                                  
    tmp_19=(((tmp_1[0])*(tmp_1[3]))-((tmp_1[2])*(tmp_1[1])));
    tmp_20[0]=((tmp_2[0])/ tmp_19);                          
    tmp_20[2]=((tmp_2[2])/ tmp_19);                          
    tmp_20[1]=((tmp_2[1])/ tmp_19);                          
    tmp_20[3]=((tmp_2[3])/ tmp_19);                          
\end{sessioncmd}

By looking at the code of function \verb!inv_bvar!, it may seem 
surprizing that its execution emits pseudo-code. But, note that affectations, 
minus operation and multiplications are also overloaded. It is precisely the role of the overloaded 
functions to produce the required pseudo-code. 

As an example of a 
function which contains explicit code directives, we give 
the overloaded code for the overloaded unary minus function:

\begin{verbatim}
function out = minus_bvar(in)
// unary minus 
  global overflow_option
  if ~is_sym(in) then out= ( - valueof(in));return,end
  if prod(size(valueof(in)))==1 then
    out=symbolics(-valueof(in),getunique())
    rhs=expression("-",list(in),overflow_option)
    gen_def(out,rhs)
  else
    out=bvarempty(in)
    sz=size(valueof(in))
    for i=1:sz(1)
      for j=1:sz(2)
        out(i,j)=-in(i,j)
      end
    end
  end
endfunction
\end{verbatim}

When the function argument is not symbolic then the function 
just performs \verb!( - valueof(in))! and no code is emitted.
When the variable is symbolic and of size \verb!1x1! the function 
returns a new symbolics variable (\verb!out=symbolics(-valueof(in),getunique()!) 
and the code is emitted by the call to the \verb!expression! function 
(\verb!expression("-",list(in),overflow_option)!) and by the call to the 
function \verb!gen_def(out,rhs)!. When the variable is a general matrix, the same function is applied to every entry of the matrix. 


\section{Code generation directives}
A number of specific Nsp functions are used as directives  to control code 
generation during the execution of the script. Some are easy to understand, for example
directives to initiate and terminate the code generation process:
\begin{verbatim}
codegen_init()
textout=codegen_finalize()
\end{verbatim}
or to start and end a function definition in the target language:
\begin{verbatim}
StartFunction(f,_io)
EndFunction()
\end{verbatim}
More complex directives are used to specify properties of the generated code that cannot be naturally
expressed in Nsp due to semantic differences between the Nsp language and the target language. The 
most important directives are described below.

\subsection{Creating persistent variables}
Persistent or static variables do not exist in Nsp but they exist in the target languages 
considered for code
generation (in particular C) and are needed. So special directives are introduced for creating them.
A pool of persistent variables may be created with the functions 
\verb+persistent_create()+ and \verb+persistent_insert(...)+. 
Each variable in the pool of persistent variables is a \verb+bvar+ object
of type symbolics. When code is emitted, a top level declaration (of static type) 
will be generated for each persistent variable used in the generated code. 
The function \verb+persistent_extract()+ can be used to obtain the persistent 
variable. 

When the function \verb+persistent_insert+ is used with a variable name 
already present in the  pool of persistent variables then a local variable 
declaration is emited and the local variable value is copied into the 
persistent variable. 

In the generated code a special function called \verb+initialize+ will 
contain code which resets all the  persistent variables to their default values. 

\begin{sessioncmd}
-nsp->// demo {\bf for} creation of a pool of persistent variables 
-nsp->codegen_init();
-nsp->states=persistent_create();
-nsp->states=persistent_insert(states,'x1',1:3);
-nsp->states=persistent_insert(states,'x2',7);
-nsp->// note that not used persistent declarations are removed from the
-nsp->// generated code (x3 is unsued).
-nsp->states=persistent_insert(states,'x3',1:6);
-nsp->io=inouts();
-nsp->StartFunction("foo",io);
-nsp->code_insert('annotation','copy [4:6] into x1 with memcpy');
-nsp->states=persistent_insert(states,'x1',[4:6]);
-nsp->code_insert('annotation','copy with assign since x2 is 1x1');
-nsp->states=persistent_insert(states,'x2',8);
-nsp->EndFunction("foo",io);
-nsp->txt=codegen_finalize() 
txt     =               s (17x1)

    static double  x1[]=\{   1,   2,   3 \};   
    static double  x2=7;                     

  void initialize()\{                         
    static double  tmp_4[]=\{   1,   2,   3 \};
    static double  tmp_5=7;                  
    memcpy(x1,tmp_4,3*sizeof(double));       
    x2=tmp_5;                                
  \}                                          

  void foo()\{                                
    double tmp_1[]=\{   4,   5,   6 \};        
    /* copy [4:6] into x1 with memcpy*/      
    memcpy(x1,tmp_1,3*sizeof(double));       
    /* copy with assign since x2 is 1x1*/    
    x2=8;                                    
  \}                                          
-nsp->txt=codegen_finalize() 
\end{sessioncmd}

\subsection{Creating function arguments}

Function arguments are created with the use of the functions 
\verb+inouts+ and \verb+inouts_insert+. Arguments are \verb+bvar+ objects 
of type symbolics. When the \verb+inouts_insert+ function is used with a variable name 
already present in a sequence of function arguments, then a local 
variable declaration is emited and the local variable value is copied into the 
function argument. 

In the following example a function \verb+foo+ with one argument is created. 
When calling the function \verb+foo+, the persistent variable \verb+x1+ is 
filled with the value of the function \verb+foo+ argument and then the 
function \verb+foo+ argument is filled with the constant value \verb+[7,8,9]+.

\begin{sessioncmd}
-nsp->codegen_init();
-nsp->states=persistent_create();
-nsp->states=persistent_insert(states,'x1',1:3);
-nsp->// in and out arguments of the {\bf function} 
-nsp->io=inouts();
-nsp->io=inouts_insert(io,'inouts1',[4,8,9]);
-nsp->// 
-nsp->StartFunction("foo",io);
-nsp->code_insert('annotation','copy inouts1 to x1 with memcpy');
-nsp->states=persistent_insert(states,'x1',io.inouts1);
-nsp->code_insert('annotation','copy [7,8,9] into inouts1');
-nsp->io = inouts_insert(io,'inouts1',[7,8,9]);
-nsp->EndFunction("foo",io);
-nsp->txt=codegen_finalize();
\end{sessioncmd}

\subsection{Function constant}

The declaration and initialization of local variables (automatic variables) may
be performed with the \verb+constant+ function.

\begin{sessioncmd}
-nsp->// declaration {\bf for} a constant
-nsp->codegen_init();
-nsp->out=constant([5,0;7,8],'x1'); // declaration + set 
-nsp->global('declarations','code','top_declarations','text');
-nsp->txt=code_printer_c(code,declarations)
txt     =               s (1x1)

    double x1[]=\{   5,   7,   0,   8 \};
\end{sessioncmd}

\subsection{Function expand}
The function \verb+expand+ returns in a new symbolic variable. In particular,
\verb+expand(in,m,n)+ returns a new symbolic \verb+bvar+ variable 
with the same type as the variable \verb+in+ but with size \verb+mxn+. 
It also inserts a declaration for the new returned variable in the generated code.

\begin{sessioncmd}
-nsp->// demo {\bf for} expand 
-nsp->codegen_init();
-nsp->in=numerics(
-nsp->var=expand(in,2,3)
var     = "tmp_2",
-nsp->global('declarations','code','top_declarations','text');
-nsp->txt=code_printer_c(code,declarations)
txt     =               s (1x1)

    int tmp_2[]=\{  TRUE,  TRUE,  TRUE,  TRUE,  TRUE,  TRUE \};
-nsp->  
-nsp->  
\end{sessioncmd}

\subsection{Functions bvarcopy and bvarempty}

The functions \verb+bvarcopy+ and \verb+bvarempty+ are used to create 
new symbolic \verb+bvar+ variables with same type and size as their 
 argument. In the case of \verb+bvarcopy+, a code for copying the 
value of the argument \verb+in+ to the value of the newly created 
variable is also emitted.

\begin{sessioncmd}
-nsp->// demo {\bf for} bvarcopy and bvarempty
-nsp->codegen_init();
-nsp->in=numerics(rand(2,3));
-nsp->var=bvarcopy(in); // var is declared and set or mcopy is generated 
-nsp->global('declarations','code','top_declarations','text');
-nsp->txt=code_printer_c(code,declarations)
txt     =               s (2x1)

 Column 1 :

    double tmp_2[]=\{   0.8147236919030547,   0.1354770041070879,   
                        0.9057919341139495,   0.8350085897836834,   
                        0.1269868118688464,   0.9688677710946649 \};
    memcpy(tmp_2,tmp_1,6*sizeof(double));                                                                                                                 
-nsp->  
-nsp->codegen_init();
-nsp->in=numerics(rand(2,3));
-nsp->var=bvarempty(in); // var is declared and filled in declaration.
-nsp->global('declarations','code','top_declarations','text');
-nsp->txt=code_printer_c(code,declarations)
txt     =               s (1x1)

 Column 1 :

    double tmp_4[]=\{   0.91337585565634072,   0.22103404277004302,   
                        0.63235924998298287,   0.30816705035977066,   
                        0.09754040162079036,   0.54722059634514153 \};
\end{sessioncmd}

\section{Application to code generation for Scicos}

\subsection{Code generation script}
Code generation script is an Nsp script generated automatically from a Scicos/VSS 
model, and in particular from
the content of the Super Block for which code generation is requested. Scicos compiler provides all the
information needed to generate this script. The information includes in particular
the list of blocks and their order of execution for the
initialization, state update and output update phases. The compilation result contains also the data types
and sizes of all the link variables inside the Super Block and those of its input and output ports.

To generate the script, the model variables are first determined. These variables contain the
signals on various links and the internal block states. They  lead to static C variables in the final C code
because they have to retain their values from one call to the next. Links that transfer constant values are partial evaluated
and do not appear in the generated code. This is done by propagating constants through the diagram, when possible.

The  script starts with a section to  determine the variables' initial values, which are then explicitly 
deefined in the generated code. This section calls the block routines with \verb+flag+~$-1$ and does not use \verb+bvar+ type
variables. Subsequently the generated script calls in proper order the block routines with flags $1$ and $2$. The input and
output values in these calls are of type \verb+bvar+ and the C code is generated thanks to operator overloading.

Unlike the code generation script, which is generated automatically and deals explicitly with variables of
type \verb+bvar+, the developer of the Nsp Scicos/VSS block functions
writes the function codes in standard Nsp as if these functions were only used for simulation. 
In fact, he does not even need to know that the 
arguments of his functions may be at any time be of any type other than regular Nsp data types. There are however certain
limitations on what he can use in his code. In particular, no conditional tests may be based on values that
cannot be constant propagated (partial evaluated to a numeric value) and the compiler should be able to determine 
the number of iterations in 
\verb+for+ and \verb+while+ loops, assuming the types and sizes of all the input arguments to his functions 
are known. Note also that not all Nsp primitives may be used; only those that have been overloaded 
for the \verb+bvar+ type may be used (most basic Nsp primitives have already been overloaded).

The constraint on the usage of conditional statements such as the \verb+if then else+ construct may seem severe, but
it turns out that for describing Scicos/VSS block behaviors, conditional statements based on non-constant propagated
conditions are rarely encountered. In most cases such conditions may be expressed as expressional 
\verb+if then else+ and \verb+select case+ statements, which are implemented as \verb+If_exp+ and \verb+Select_exp+ functions.

An expressional \verb+if then else+ statement expressed as \verb+If_exp+ function is used as follows:
\begin{verbatim}
out = If_exp(cond, exp1, exp2)
\end{verbatim}
If \verb+cond+ is true, \verb+out+ takes the value of the expression \verb+exp1+, if not, it takes the value 
of the expression \verb+exp2+. Note however that in either case both expressions \verb+exp1+ and \verb+exp2+ are
evaluated, unlike for a real \verb+if then else+ statement where only one of the expressions is evaluated. These
structurel conditions used in particular for subsampling in Scicos and VSS are realized using the special blocks 
\verb+IfThenElse+ and \verb+ESelect+.

\subsection{Conditional subsampling}
We have seen so far that in order for the code generation script to function, the expressions of all
conditional statements (\verb!if-then-else!, \verb!when!,...) must be partial evaluated, which means that
it cannot be dependent on symbolic variables. But Scicos and VSS  models may include conditional subsampling using
in particular the \verb!IfThenElse! block. Subsampling is an important mechanism available in Scicos, so a special
method is introduced to handle such models for efficient code generation.

\begin{figure}[htb]
  \begin{center}
    \mybox{\includegraphics[scale=\scicosscale]{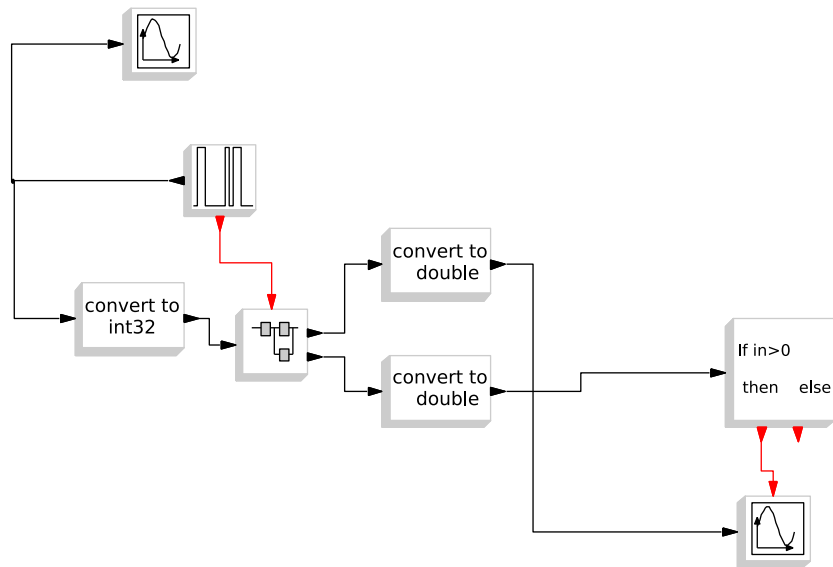}}
    \caption{Model for testing the coding scheme.\label{code}}
  \end{center}
\end{figure}

\begin{figure}[htb]
  \begin{center}
    \mybox{\includegraphics[scale=\scicosscale]{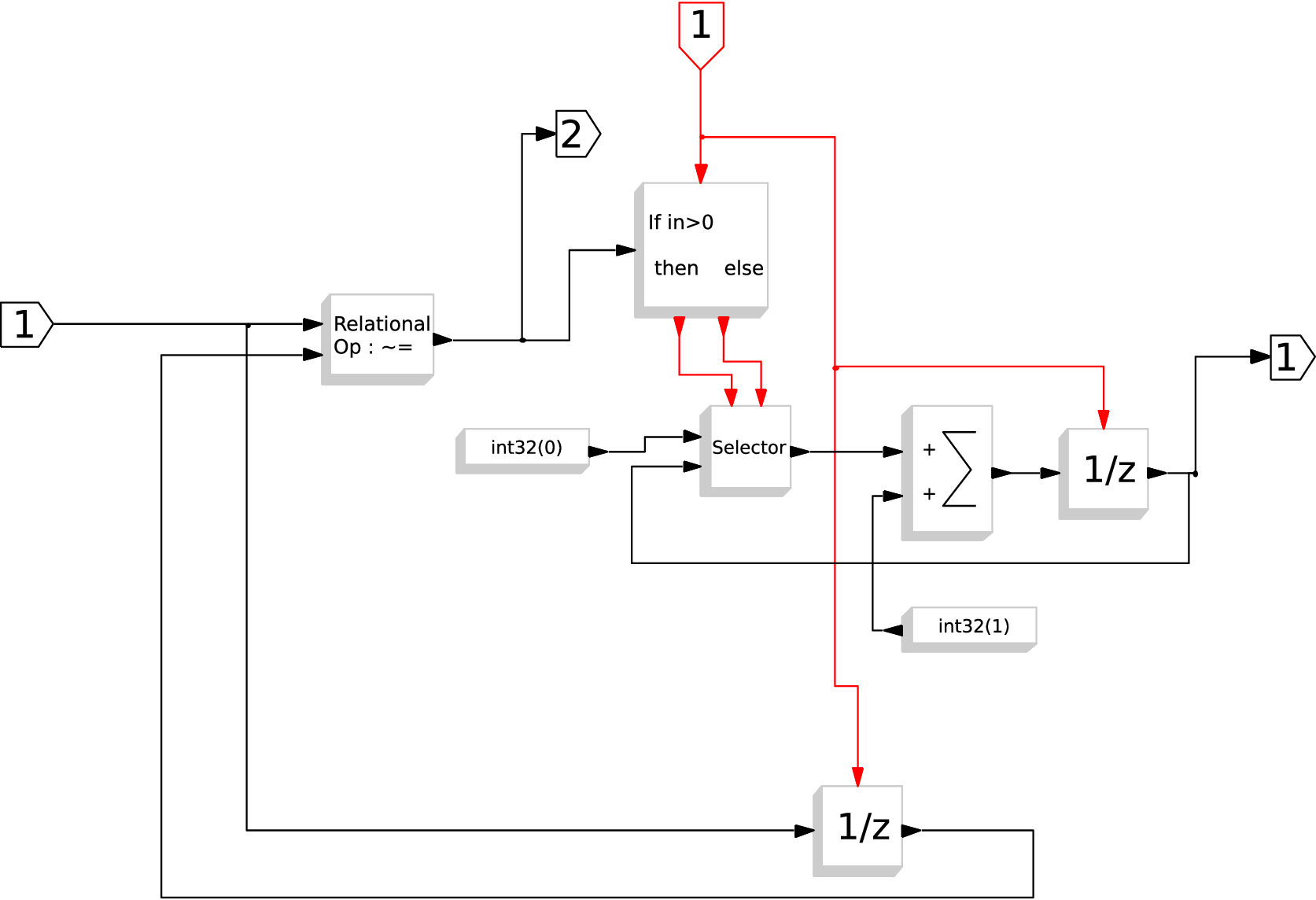}}
    \caption{The Super Block modeling the coding scheme using subsampling. 
Code is generated for this Super Block. \label{code_SB}}
  \end{center}
\end{figure}

An example of Scicos model with conditional subsampling is given in Fig.~\ref{code}. The content of the Super Block
for which C code is generated is illustrated in Fig.~\ref{code_SB}. This model implements a simple
coding scheme where the Super Block receives a sequence of $0$ and $1$ integers and generates a sequence providing the
number of consecutive $0$'s or $1$'s. The Super Block contains an  \verb+IfThenElse+ block for testing whether or not
two successive integers have the the same value, and depending on the result, a counter is either incremented or reset to~$1$.

\clearpage

The generated code shows that the test implemented by the \verb+IfThenElse+ block has produced an \verb+if+ statement in~C:
\begin{verbatim}
/* Scicos Computational function  
 * Generated by Code_Generation toolbox of Scicos with scicos4.4.1
 * date : 05 mars 2015
 */
#include <scicos/scicos_block4.h>
#include <string.h>
#include <stdio.h>
#include <stdlib.h>
#include <stdint.h>
#include <math.h>
/* Start1004*/

  static int32_t  z_10041=0;
  static int32_t  z_10042=0;
  static int32_t  link10046=0;
  static int32_t  link10048=1;

void initialize1004(){
  static int32_t  tmp_13=0;
  static int32_t  tmp_14=0;
  static int32_t  tmp_15=0;
  static int32_t  tmp_16=1;
  z_10041=tmp_13;
  z_10042=tmp_14;
  link10046=tmp_15;
  link10048=tmp_16;
}

void updateOutput10041(int32_t *inouts1,int32_t *inouts2,int32_t *inouts3){
  
  /* Selct block starts*/
  /* Selct block ends*/
  link10046=0;
}

void updateOutput10042(int32_t *inouts1,int32_t *inouts2,int32_t *inouts3){
  
  /* Selct block starts*/
  /* Selct block ends*/
  link10046=*inouts2;
}

void updateOutput10043(int32_t *inouts1,int32_t *inouts2,int32_t *inouts3){
  int tmp_6;
  int tmp_7;
  /* RELATIONALOP block starts*/
  tmp_6=(*inouts1!=z_10041);
  /* RELATIONALOP block ends*/
  *inouts3=tmp_6;
  *inouts2=z_10042;
  tmp_7=(*inouts3>0);
  if (tmp_7) {
    updateOutput10041(inouts1,inouts2,inouts3);
  } else {
    updateOutput10042(inouts1,inouts2,inouts3);
  }
  /* Sum block begins with 2 inputs.*/
  link10048=(link10046+1);
}

void updateState10043(int32_t *inouts1,int32_t *inouts2,int32_t *inouts3){
  int tmp_11;
  z_10041=*inouts1;
  /* RELATIONALOP block starts*/
  /* RELATIONALOP block ends*/
  z_10042=link10048;
  tmp_11=(*inouts3>0);
}
/* End1004*/

void toto1004(scicos_block *block,int flag)
  {
  if (flag == 1) {
   updateOutput10043((Getint32InPortPtrs(block,1)),(Getint32OutPortPtrs(block,1)),(Getint32OutPortPtrs(block,2)));
  }
  else if (flag == 2) {
   updateState10043((Getint32InPortPtrs(block,1)),(Getint32OutPortPtrs(block,1)),(Getint32OutPortPtrs(block,2)));
  }
  else if (flag == 4) {
     initialize1004();
  }
}
\end{verbatim}

The code geneation script uses the \verb+if_cos+ function, which uses the \verb+If_exp+ function.
The \verb+if_cos+ function is defined as follows:
\begin{verbatim}
function if_cos(in,f1,f2)
// f1 and f2 are expressions of type call(..)
  if is_sym(in) then 
    code_insert("if_expr", in > (0),f1,f2);
  elseif valueof(in)>0 then
    code_insert("ident",f1)   
  else
    code_insert("ident",f2)   
  end
endfunction
\end{verbatim}
Note that the \verb+if_cos+ function does not return any value. This is consistent with the fact that the
\verb+IfThenElse+ block has no regular output. The second and third arguments of this function are calls to external
(generated) C routines. The arguments are the Super Block inputs and outputs, passed as pointers, so that these
external routines can only modify these inputs/outputs (in general only outputs) and global variables.

\subsection{Example}
The methodology presented in this paper may be used to generate code for models using existing Scicos blocks
but it can also be used to generate code for models containing \verb+sciblk+s, i.e. blocks where the user has
expressed the block's behavior in Nsp. The following example is adapted from an extended Kalman filter
application developed in \cite{god} for Simulink with an embedded Matlab block.
\begin{figure}[ht]
  \begin{center}
    \mybox{\includegraphics[scale=\scicosscale]{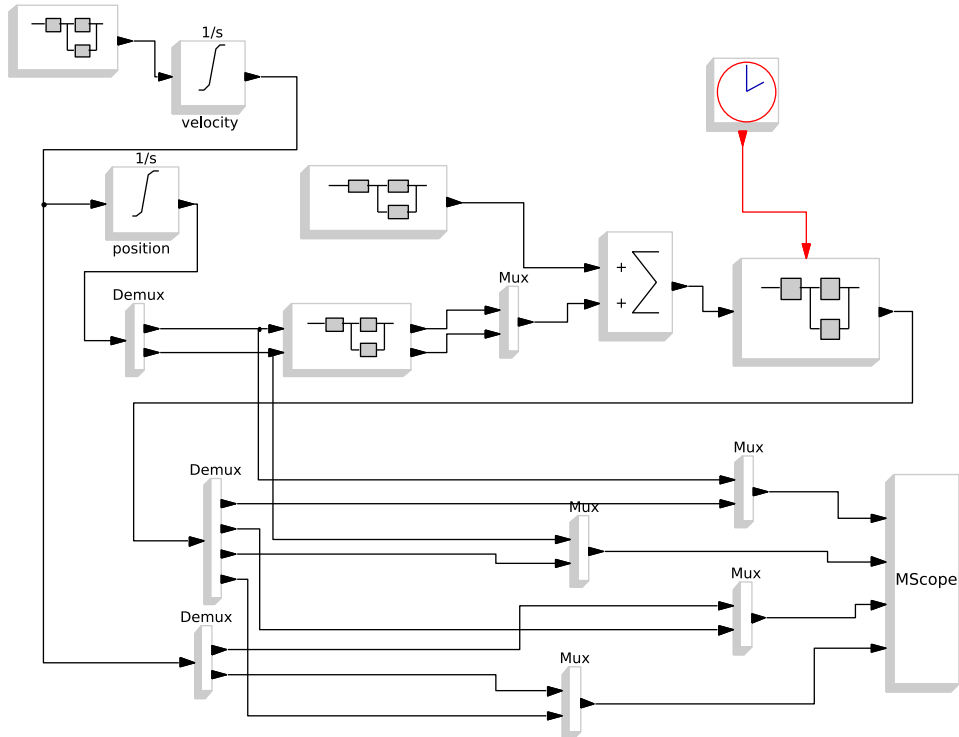}}
    \caption{Model of an extended Kalman filter.\label{k1}}
  \end{center}
\end{figure}

\begin{figure}[ht]
  \begin{center}
    \mybox{\includegraphics[scale=\scicosscale]{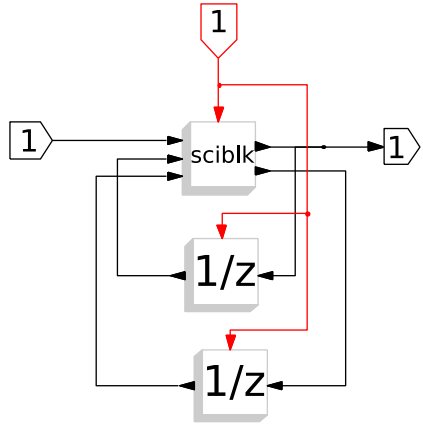}}
    \caption{The Nsp  block is included in a Super Block for code generation. The delay blocks are used to 
hold the state of the filter and the corresponding covariance matrix.\label{k2}}
  \end{center}
\end{figure}

The Kalman filter equation is expressed in Nsp in the \verb+sciblk+ block:
\begin{verbatim}
function [blk] = P_sciblk(blk,flag)

if flag==1 then  //output computation
meas=blk.io(1);xhat=blk.io(2);P=blk.io(3);

//put_annotation("sciblk block begins.")

// This Nsp Function implements an extended Kalman filter used
// for object tracking.
//
// The states of the process are given by
// x = [x_position; x_velocity; y_position; y_velocity];
//
// and the measurements are given by
// y = [range; bearing]
//
// where
// range = sqrt(x_position^2 + y_position^2)
// bearing = atan(y_position/x_position)

// Author: Phil Goddard (phil@goddardconsulting.ca)
// Date: Q2, 2011.
// Adapted to Nsp/Scicos in 2015
	dt=0.1
Q = diag([0 .1 0 .1]);
R = diag([50^2 0.005^2]);
// Calculate the Jacobians for the state and measurement equations
F = [1 dt 0 0;0 1 0 0;0 0 1 dt;0 0 0 1];
rangeHat = sqrt(xhat(1)^2+xhat(3)^2);
bearingHat = atan(xhat(3),xhat(1));
yhat = [rangeHat; bearingHat];
H = [cos(bearingHat) 0 sin(bearingHat) 0;
-sin(bearingHat)/rangeHat 0 cos(bearingHat)/rangeHat 0];
// Propogate the state and covariance matrices
xhat = F*xhat;
P = F*P*F' + Q;
// Calculate the Kalman gain
K = P*H'/(H*P*H' + R);
// Calculate the measurement residual
resid = meas - yhat;
// Update the state and covariance estimates
xhat = xhat + K*resid;
P = (eye(size(K,1),size(K,1))-K*H)*P;
// Post the results
blk.io(4)=xhat
blk.io(5)=P
elseif flag==2 then  //discrete state computation (if any)

end
//put_annotation("sciblk block ends.")
endfunction
\end{verbatim}
Note that this is a very general Nsp code; it uses a number of matrix manipulation operators and functions. All these operators
and functions are overloaded. The constraints on the usage
of conditional statements and loops are also satisfied since there are no such statements in the code.

\section{Conclusion}
This paper has provided a short presentation of a new methodology based on partial evaluation and overloading 
used for the implementation of a code generation tool developed for Scicos and VSS. 
The code is freely available in Nsp 1.0 (See \cite{nsp}).

\appendix
\section{C code generated for the extended Kalman filter example}

\begin{Verbatim}[fontsize=\tiny]
/* Scicos Computational function  
 * Generated by Code_Generation toolbox of Scicos with scicos4.4.1
 * date : 06 mars 2015
 */
#include <scicos/scicos_block4.h>
#include <string.h>
#include <stdio.h>
#include <stdlib.h>
#include <stdint.h>
#include <math.h>
/* Start1002*/
void quote(double *res, double *a, double *dm,double *dn)
{
 int i,j, m1=(int) (*dm), n1 = (int) (*dn) ;
 for (i = 0 ; i < (m1); i++) 
   for (j = 0 ; j < (n1); j++) 
     {
       res[j+(n1)*i]= a[i+(m1)*j];
     }
}

void mult(double *res, double *a, double *b,double *md1,double *nd1,double *md2,double *nd2)
{
 int i,j,k,m1=(int) (*md1),n1= (int) (*nd1),m2= (int) (*md2),n2=(int) (*nd2);
 for (i = 0 ; i < m1; i++) 
   for (j = 0 ; j < n2; j++) 
     {
       res[i+m1*j]=0;
       for (k = 0 ; k < n1; k++) 
         res[i+(m1)*j] += a[i+(m1)*k]*b[k+(m2)*j];
     }
}


  static double  z_10021[]={   -900,     80,    950,     20 };
  static double  z_10022[]={   0,   0,   0,   0,   0,   0,   0,   0,   0,   0,   0,   0,   0,   0,   0,   0 };
  static double  link10024[]={   0,   0,   0,   0,   0,   0,   0,   0,   0,   0,   0,   0,   0,   0,   0,   0 };

void initialize1002(){
  static double  tmp_400[]={   -900,     80,    950,     20 };
  static double  tmp_401[]={   0,   0,   0,   0,   0,   0,   0,   0,   0,   0,   0,   0,   0,   0,   0,   0 };
  static double  tmp_402[]={   0,   0,   0,   0,   0,   0,   0,   0,   0,   0,   0,   0,   0,   0,   0,   0 };
  memcpy(z_10021,tmp_400,4*sizeof(double));
  memcpy(z_10022,tmp_401,16*sizeof(double));
  memcpy(link10024,tmp_402,16*sizeof(double));
}

void updateOutput10021(double *inouts1,double *inouts2){
  double tmp_3;
  double tmp_5;
  double tmp_8;
  double tmp_11;
  double tmp_12[2];
  double tmp_18[2];
  double tmp_23[3];
  double tmp_30[4];
  double tmp_41[2];
  double tmp_47[3];
  double tmp_54[4];
  double tmp_62[8];
  double tmp_79[4];
  double tmp_104[16];
  double tmp_105[]={   1.0000000000000000,                    0 ,
                     0,                    0 ,
    0.1000000000000000,   1.0000000000000000 ,
                     0,                    0 ,
                     0,                    0 ,
    1.0000000000000000,                    0 ,
                     0,                    0 ,
    0.1000000000000000,   1.0000000000000000 };
  double tmp_106=4;
  double tmp_107=4;
  double tmp_108=4;
  double tmp_109=4;
  double tmp_110[16];
  double tmp_111[]={   1.0000000000000000,   0.1000000000000000 ,
                     0,                    0 ,
                     0,   1.0000000000000000 ,
                     0,                    0 ,
                     0,                    0 ,
    1.0000000000000000,   0.1000000000000000 ,
                     0,                    0 ,
                     0,   1.0000000000000000 };
  double tmp_112=4;
  double tmp_113=4;
  double tmp_114=4;
  double tmp_115=4;
  double tmp_116[16];
  double tmp_151[8];
  double tmp_152=2;
  double tmp_153=4;
  double tmp_154[8];
  double tmp_155=4;
  double tmp_156=4;
  double tmp_157=4;
  double tmp_158=2;
  double tmp_159[8];
  double tmp_160=2;
  double tmp_161=4;
  double tmp_162=4;
  double tmp_163=4;
  double tmp_164[8];
  double tmp_165=2;
  double tmp_166=4;
  double tmp_167[4];
  double tmp_232[4];
  double tmp_243[4];
  double tmp_260;
  double tmp_261[4];
  double tmp_274[8];
  double tmp_275=4;
  double tmp_276=2;
  double tmp_277=2;
  double tmp_278=2;
  double tmp_279[2];
  double tmp_288[4];
  double tmp_321[4];
  double tmp_338[16];
  double tmp_339=4;
  double tmp_340=2;
  double tmp_341=2;
  double tmp_342=4;
  double tmp_343[16];
  double tmp_392[16];
  double tmp_393=4;
  double tmp_394=4;
  double tmp_395=4;
  double tmp_396=4;
  tmp_3=(z_10021[0]);
  tmp_5=(z_10021[2]);
  tmp_8=sqrt(((tmp_3*tmp_3)+(tmp_5*tmp_5)));
  tmp_11=atan2((z_10021[2]),(z_10021[0]));
  tmp_12[0]=tmp_8;
  tmp_12[1]=tmp_11;
  /* Begin concatr of tmp_17 with unknown*/
  tmp_18[0]=cos(tmp_11);
  tmp_18[1]=0;
  /* end concatr of tmp_17 with unknown*/
  /* Begin concatr of tmp_21 with tmp_22*/
  tmp_23[0]=(tmp_18[0]);
  tmp_23[1]=(tmp_18[1]);
  tmp_23[2]=sin(tmp_11);
  /* end concatr of tmp_21 with tmp_22*/
  /* Begin concatr of tmp_29 with unknown*/
  tmp_30[0]=(tmp_23[0]);
  tmp_30[1]=(tmp_23[1]);
  tmp_30[2]=(tmp_23[2]);
  tmp_30[3]=0;
  /* end concatr of tmp_29 with unknown*/
  /* Begin concatr of tmp_40 with unknown*/
  tmp_41[0]=(-(sin(tmp_11)/ tmp_8));
  tmp_41[1]=0;
  /* end concatr of tmp_40 with unknown*/
  /* Begin concatr of tmp_44 with tmp_46*/
  tmp_47[0]=(tmp_41[0]);
  tmp_47[1]=(tmp_41[1]);
  tmp_47[2]=(cos(tmp_11)/ tmp_8);
  /* end concatr of tmp_44 with tmp_46*/
  /* Begin concatr of tmp_53 with unknown*/
  tmp_54[0]=(tmp_47[0]);
  tmp_54[1]=(tmp_47[1]);
  tmp_54[2]=(tmp_47[2]);
  tmp_54[3]=0;
  /* end concatr of tmp_53 with unknown*/
  tmp_62[0]=(tmp_30[0]);
  tmp_62[1]=(tmp_54[0]);
  tmp_62[2]=(tmp_30[1]);
  tmp_62[3]=(tmp_54[1]);
  tmp_62[4]=(tmp_30[2]);
  tmp_62[5]=(tmp_54[2]);
  tmp_62[6]=(tmp_30[3]);
  tmp_62[7]=(tmp_54[3]);
  tmp_79[0]=((z_10021[0])+(0.1*(z_10021[1])));
  tmp_79[1]=(z_10021[1]);
  tmp_79[2]=((z_10021[2])+(0.1*(z_10021[3])));
  tmp_79[3]=(z_10021[3]);
  /* Product of matrices resulting size 16>6: calling external function*/
  tmp_105[0]=1;
  tmp_105[1]=0;
  tmp_105[2]=0;
  tmp_105[3]=0;
  tmp_105[4]=0.1;
  tmp_105[5]=1;
  tmp_105[6]=0;
  tmp_105[7]=0;
  tmp_105[8]=0;
  tmp_105[9]=0;
  tmp_105[10]=1;
  tmp_105[11]=0;
  tmp_105[12]=0;
  tmp_105[13]=0;
  tmp_105[14]=0.1;
  tmp_105[15]=1;
  tmp_106=4;
  tmp_107=4;
  tmp_108=4;
  tmp_109=4;
mult(tmp_104,tmp_105,z_10022,&tmp_106,&tmp_107,&tmp_108,&tmp_109);
  /* Product of matrices resulting size 16>6: calling external function*/
  tmp_111[0]=1;
  tmp_111[1]=0.1;
  tmp_111[2]=0;
  tmp_111[3]=0;
  tmp_111[4]=0;
  tmp_111[5]=1;
  tmp_111[6]=0;
  tmp_111[7]=0;
  tmp_111[8]=0;
  tmp_111[9]=0;
  tmp_111[10]=1;
  tmp_111[11]=0.1;
  tmp_111[12]=0;
  tmp_111[13]=0;
  tmp_111[14]=0;
  tmp_111[15]=1;
  tmp_112=4;
  tmp_113=4;
  tmp_114=4;
  tmp_115=4;
mult(tmp_110,tmp_104,tmp_111,&tmp_112,&tmp_113,&tmp_114,&tmp_115);
  tmp_116[0]=(tmp_110[0]);
  tmp_116[4]=(tmp_110[4]);
  tmp_116[8]=(tmp_110[8]);
  tmp_116[12]=(tmp_110[12]);
  tmp_116[1]=(tmp_110[1]);
  tmp_116[5]=((tmp_110[5])+0.1);
  tmp_116[9]=(tmp_110[9]);
  tmp_116[13]=(tmp_110[13]);
  tmp_116[2]=(tmp_110[2]);
  tmp_116[6]=(tmp_110[6]);
  tmp_116[10]=(tmp_110[10]);
  tmp_116[14]=(tmp_110[14]);
  tmp_116[3]=(tmp_110[3]);
  tmp_116[7]=(tmp_110[7]);
  tmp_116[11]=(tmp_110[11]);
  tmp_116[15]=((tmp_110[15])+0.1);
  /* Transpose of matrix of size 8>6: calling external function*/
  tmp_152=2;
  tmp_153=4;
quote(tmp_151,tmp_62,&tmp_152,&tmp_153);
  /* End of Transpose*/
  /* Product of matrices resulting size 8>6: calling external function*/
  tmp_155=4;
  tmp_156=4;
  tmp_157=4;
  tmp_158=2;
mult(tmp_154,tmp_116,tmp_151,&tmp_155,&tmp_156,&tmp_157,&tmp_158);
  /* Product of matrices resulting size 8>6: calling external function*/
  tmp_160=2;
  tmp_161=4;
  tmp_162=4;
  tmp_163=4;
mult(tmp_159,tmp_62,tmp_116,&tmp_160,&tmp_161,&tmp_162,&tmp_163);
  /* Transpose of matrix of size 8>6: calling external function*/
  tmp_165=2;
  tmp_166=4;
quote(tmp_164,tmp_62,&tmp_165,&tmp_166);
  /* End of Transpose*/
  tmp_167[0]=(((((tmp_159[0])*(tmp_164[0]))+((tmp_159[2])*(tmp_164[1])))+((tmp_159[4])*(tmp_164[2])))+((tmp_159[6])*(tmp_164[3])));
  tmp_167[2]=(((((tmp_159[0])*(tmp_164[4]))+((tmp_159[2])*(tmp_164[5])))+((tmp_159[4])*(tmp_164[6])))+((tmp_159[6])*(tmp_164[7])));
  tmp_167[1]=(((((tmp_159[1])*(tmp_164[0]))+((tmp_159[3])*(tmp_164[1])))+((tmp_159[5])*(tmp_164[2])))+((tmp_159[7])*(tmp_164[3])));
  tmp_167[3]=(((((tmp_159[1])*(tmp_164[4]))+((tmp_159[3])*(tmp_164[5])))+((tmp_159[5])*(tmp_164[6])))+((tmp_159[7])*(tmp_164[7])));
  tmp_232[0]=((tmp_167[0])+2500);
  tmp_232[2]=(tmp_167[2]);
  tmp_232[1]=(tmp_167[1]);
  tmp_232[3]=((tmp_167[3])+0.000025);
  tmp_243[0]=(tmp_232[3]);
  tmp_243[3]=(tmp_232[0]);
  tmp_243[2]=(-(tmp_232[2]));
  tmp_243[1]=(-(tmp_232[1]));
  tmp_260=(((tmp_232[0])*(tmp_232[3]))-((tmp_232[2])*(tmp_232[1])));
  tmp_261[0]=((tmp_243[0])/ tmp_260);
  tmp_261[2]=((tmp_243[2])/ tmp_260);
  tmp_261[1]=((tmp_243[1])/ tmp_260);
  tmp_261[3]=((tmp_243[3])/ tmp_260);
  /* Product of matrices resulting size 8>6: calling external function*/
  tmp_275=4;
  tmp_276=2;
  tmp_277=2;
  tmp_278=2;
mult(tmp_274,tmp_154,tmp_261,&tmp_275,&tmp_276,&tmp_277,&tmp_278);
  tmp_279[0]=((inouts1[0])-(tmp_12[0]));
  tmp_279[1]=((inouts1[1])-(tmp_12[1]));
  tmp_288[0]=(((tmp_274[0])*(tmp_279[0]))+((tmp_274[4])*(tmp_279[1])));
  tmp_288[1]=(((tmp_274[1])*(tmp_279[0]))+((tmp_274[5])*(tmp_279[1])));
  tmp_288[2]=(((tmp_274[2])*(tmp_279[0]))+((tmp_274[6])*(tmp_279[1])));
  tmp_288[3]=(((tmp_274[3])*(tmp_279[0]))+((tmp_274[7])*(tmp_279[1])));
  tmp_321[0]=((tmp_79[0])+(tmp_288[0]));
  tmp_321[1]=((tmp_79[1])+(tmp_288[1]));
  tmp_321[2]=((tmp_79[2])+(tmp_288[2]));
  tmp_321[3]=((tmp_79[3])+(tmp_288[3]));
  /* Product of matrices resulting size 16>6: calling external function*/
  tmp_339=4;
  tmp_340=2;
  tmp_341=2;
  tmp_342=4;
mult(tmp_338,tmp_274,tmp_62,&tmp_339,&tmp_340,&tmp_341,&tmp_342);
  tmp_343[0]=(1-(tmp_338[0]));
  tmp_343[4]=(-(tmp_338[4]));
  tmp_343[8]=(-(tmp_338[8]));
  tmp_343[12]=(-(tmp_338[12]));
  tmp_343[1]=(-(tmp_338[1]));
  tmp_343[5]=(1-(tmp_338[5]));
  tmp_343[9]=(-(tmp_338[9]));
  tmp_343[13]=(-(tmp_338[13]));
  tmp_343[2]=(-(tmp_338[2]));
  tmp_343[6]=(-(tmp_338[6]));
  tmp_343[10]=(1-(tmp_338[10]));
  tmp_343[14]=(-(tmp_338[14]));
  tmp_343[3]=(-(tmp_338[3]));
  tmp_343[7]=(-(tmp_338[7]));
  tmp_343[11]=(-(tmp_338[11]));
  tmp_343[15]=(1-(tmp_338[15]));
  /* Product of matrices resulting size 16>6: calling external function*/
  tmp_393=4;
  tmp_394=4;
  tmp_395=4;
  tmp_396=4;
mult(tmp_392,tmp_343,tmp_116,&tmp_393,&tmp_394,&tmp_395,&tmp_396);
  memcpy(link10024,tmp_392,16*sizeof(double));
  memcpy(inouts2,tmp_321,4*sizeof(double));
}

void updateState10021(double *inouts1,double *inouts2){
  
  memcpy(z_10021,inouts2,4*sizeof(double));
  memcpy(z_10022,link10024,16*sizeof(double));
}
/* End1002*/
void toto1002(scicos_block *block,int flag)
  {
  if (flag == 1) {
   updateOutput10021((GetRealInPortPtrs(block,1)),(GetRealOutPortPtrs(block,1)));
  }
  else if (flag == 2) {
   updateState10021((GetRealInPortPtrs(block,1)),(GetRealOutPortPtrs(block,1)));
  }
  else if (flag == 4) {
     initialize1002();
  }
}
\end{Verbatim}

\end{document}